\documentclass[12pt]{article}

\usepackage{latexsym,amsmath,amsthm,amssymb,amsfonts,amsbsy,calrsfs,multibox}
\usepackage{graphicx}

\newtheorem*{theorem}{Theorem}

\def\a{\alpha}
\def\b{\beta}
\def\g{\gamma}
\def\c{\gamma}
\def\d{\delta}

\def\l{\lambda}

\def\m{\mu}
\def\n{\nu}

\def\r{\rho}
\def\s{\sigma}

\def\t{\tau}

\def\be{\begin{equation}}
\def\ee{\end{equation}}
\def\bea{\begin{eqnarray}}
\def\eea{\end{eqnarray}}
\def\beq{\begin{eqnarray}}
\def\eeq{\end{eqnarray}}
\def\bs{{\textbf{s}}}

\def\ba{\begin{array}}
\def\ba{\end{array}}

\begin{document} 

\thispagestyle{empty}

\vspace{.6cm}

\begin{centering}

{\Large {\bf Strong obstruction of\\
  \vspace*{.2cm}
  the Berends--Burgers--van Dam spin-3 vertex}}

\vspace{2cm}

\begin{center}
{\large Xavier Bekaert$^a$, Nicolas Boulanger$^b$ and Serge Leclercq$^b$}\\
\vspace{1cm}
{\small $^a$ Laboratoire de Math\'ematiques et Physique Th\'eorique}\\
{\small Unit\'e Mixte de Recherche $6083$ du CNRS}\\
{\small F\'ed\'eration de Recherche $2964$ Denis Poisson}\\
{\small Universit\'e Fran%\c
{c}ois Rabelais, Parc de Grandmount}\\
{\small 37200 Tours, France} \\
\vspace{1mm}{\tt \footnotesize Xavier.Bekaert@lmpt.univ-tours.fr}\\
\vspace{4mm}{\small $^b$ Service de M\'ecanique et Gravitation}\\ 
{\small Universit\'e de Mons -- UMons, 20 Place du Parc}\\
{\small 7000 Mons, Belgium}\\
\vspace{1mm}{\tt \footnotesize nicolas.boulanger@umons.ac.be}\\
\vspace{1mm}{\tt \footnotesize serge.leclercq@umons.ac.be}\\

\end{center}

\end{centering}

\vspace*{.5cm}

\begin{abstract}
In the eighties, Berends, Burgers and van Dam (BBvD) found a nonabelian cubic vertex for self-interacting massless fields of spin three in flat spacetime. However, they also found that this deformation is inconsistent at higher order for any multiplet of spin-three fields. 
For arbitrary symmetric gauge fields, we severely constrain the possible nonabelian deformations of the gauge algebra and, using these results, prove that the BBvD obstruction cannot be cured by any means, even by introducing fields of spin higher (or lower) than three.
\end{abstract}

\vspace*{.2cm}

\pagebreak

\setcounter{page}{1}

\section{Introduction \label{sec:Introduction}}

One of the main features of higher spin theories is that apparently the only way of building a nonabelian interacting theory is to consider an infinite set of fields with unbounded 
value of the spin. Such an interacting theory is by now well-known: Vasiliev's equations \cite{Vasiliev:1990en,Vasiliev:1992av,Vasiliev:2003ev}. 
These equations admit (anti) de Sitter spacetime $(A)dS$ as exact solution, 
but not the flat limit case with vanishing cosmological constant.
The theory \cite{Vasiliev:1990en,Vasiliev:1992av} can be given a Lagrangian formulation \cite{BS1}, albeit of a non-standard type. 
It has been proved recently \cite{Boulanger:2008tg} that an interacting nonabelian theory built as a perturbative deformation of the free Fronsdal theory in $(A)dS$ spacetime \cite{Fronsdal:1979vb} and containing
an infinite tower of totally symmetric tensor gauge fields with unbounded 
spin, does not admit any consistent flat limit. There are thus doubts about the mere existence of a nonabelian theory around Minkoswki spacetime.

To emphasize this, we show in this paper that a standard requirement about 
higher spin nonabelian interactions in Minkowski spacetime cannot, at least
in dimension four, solve the usual flat-space interaction problems. 
Some years ago, Berends, Burgers and van Dam (BBvD) exhibited a pure spin-3 nonabelian cubic vertex in flat spacetime \cite{Berends:1984wp}, and then found that this vertex cannot be further extended to higher orders in deformations 
if only spin-3 fields are considered in the spectrum of fields \cite{Berends:1984rq}. A similar result had also been obtained in \cite{Bengtsson:1983bp}. 

In more technical terms, 
an obstruction to the existence of pure spin-3 quartic deformations appeared. 
In the light of this negative result, they brought the idea that the obstruction, as well as others of the same kind, could probably be cured by introducing fields of spin higher than 3. Recursively, this would suggest 
that every value of the spin is needed in order to build a nonabelian higher spin theory in flat spacetime. 

In this paper, we actually prove that the BBvD vertex is in fact strongly obstructed in dimension strictly higher than three, in the following sense:  
Even upon introducing higher (and/or lower) spin gauge fields, it is not possible to cure the obstruction brought in by the spin-3 BBvD vertex found 
in \cite{Berends:1984wp}. 
The very reason for this strong obstruction in flat background is that, as opposed to what happens in $(A)dS$ background, the number of derivatives involved in an expression constitutes a well-defined grading that increases with the value of the spin. 
In $(A)dS$ instead, the non-zero commutators of covariant derivatives introduce expansions in powers of the cosmological constant involving different numbers of derivatives. Furthermore, these expansions in powers of the cosmological constant are precisely what prevents one from considering a consistent nonabelian flat limit of the $(A)dS$ theories, as was mentioned
already in \cite{Fradkin:1987ks,Fradkin:1987qy}. 

The paper is organized as follows. In Section \ref{sec:af} we review Fronsdal's theory in the antifield formulation, as well as the cohomological reformulation of the consistent deformation problem. 
In Section \ref{sec:genr} we present our theorem on the lowest-order deformations of the gauge algebra arising from local cubic interactions between totally symmetric higher spin gauge fields in flat spacetime. This is used in Section \ref{sec:obstr} in order to address the specific case of the BBvD vertex. In Section \ref{concl}, these results are briefly summarized as a conclusion.

%**************************************************************************
\section{Antifield formulation and consistent deformations  \label{sec:af}}
%**************************************************************************

We use the antifield formalism \cite{Henneaux:1992ig,Barnich:1993vg,Henneaux:1998i,Barnich:1995db} to derive our results about consistent deformations of Fronsdal Lagrangians \cite{Fronsdal:1978rb}. The BBvD vertex only involves the spin-3 gauge fields, but since we allow them to mix with other gauge fields, we recall the Fronsdal Lagrangian for an arbitrary spin gauge field in flat spacetime, as well as the corresponding antifield formulation.

%-------------------------------------
\subsection{The Fronsdal Lagrangian}
%-------------------------------------

The field denoted by $\phi^a_{\m_1...\m_{s_a}}$ is a totally symmetric field of spin $s_a$ and double-traceless : $\eta^{\m_1\m_2}\eta^{\m_3\m_4}\phi^a_{\m_1...\m_{s_a}}\equiv 0\,$. 
The index $a$ labels a given set of fields with various values of the spin. 
The Fronsdal tensor reads: 
\begin{eqnarray}
F^a_{\m_1...\m_{s_a}} := \Box \phi^a_{\m_1...\m_{s_a}}-s_a\,\partial^{\r}\partial^{}_{(\m_1}\phi^a_{\m_2...\m_{s_a})\r}+\frac{s_a(s_a-1)}{2}\;\partial^{}_{(\m_1}\partial^{}_{\m_2}\phi'{}^a_{\m_3...\m_{s_a})}\quad,
\end{eqnarray} 
where $\phi'{}^a$ stands for the trace of $\phi^a\,$. 
The Fronsdal tensor is invariant under the gauge transformations:
\begin{eqnarray}
\d_\xi\phi^a_{\m_1...\m_{s_a}}=s_a\,\partial^{}_{(\m_1}\xi^a_{\m_2...\m_{s_a})}\quad,\label{gts}
\end{eqnarray} 
where the gauge parameter $\xi^a$ is traceless. 
The generalized Einstein tensor is defined as:
\begin{eqnarray}
G^a_{\m_1...\m_{s_a}} := F^a_{\m_1...\m_{s_a}} - \frac{s_a(s_a-1)}{4}\;\eta^{}_{(\m_1\m_2}F'{}^a_{\m_3...\m_{s_a})}
\quad.
\end{eqnarray} 
Finally, the Lagrangian can be written: \begin{eqnarray}{\cal{L}}_F=\frac{1}{2}\sum_a \phi^{\m_1...\m_{s_a}}_a G^a_{\m_1...\m_{s_a}}\quad.\end{eqnarray}

\subsection{Antifield formulation}

A set of fermionic ghosts $C_{\m_1...\m_{s_a-1}}^a$ is introduced, with the same tensorial 
structure as the associated gauge parameter. In particular, the ghosts are traceless.
They carry a pure ghost number 1 (also denoted $pgh\ C^a=1$). Then, two families of antifields are associated with the fields and the ghosts : the fermionic antifields $\phi^{*\m_1...\m_{s_a}}_a$ and the bosonic antifields $C^{*\m_1...\m_{s_a-1}}_a$. The antifield number (denoted $antigh$) counts the number of antifields with the following weight: $antigh\ \phi^*_a=1$ and $antigh\ C^*_a=2$.

The longitudinal derivative $\g$ has a vanishing action on every field except $\phi^a$, for which: 
\begin{eqnarray}
\g \phi^a_{\m_1...\m_{s_a}} = s_a\, \partial^{}_{(\m_1}C^a_{\m_2...\m_{s_a})}\quad.
\end{eqnarray} 
On the other hand the Kozsul--Tate differential has a non-vanishing action only on the antifields: \begin{eqnarray}
\d\phi^{*\m_1...\m_{s_a}}_a &:=& \frac{\d {\cal{L}}_F}{\d \phi^a_{\m_1...\m_{s_a}}}
~~\textrm{ and }
\nonumber \\
\d C^{*\m_1...\m_{s_a-1}}_a &:=&
-s_a\,\partial^{}_\r\left[\phi^{*\m_1...\m_{s_a-1}\r}_a-\frac{(s-1)(s-2)}{n+2s-6}\;\eta^{(\m_1\m_2}
\phi_a^*{}'{}^{\m_3...\m_{s_a-1})\r}\right]\quad ,
\end{eqnarray}
where $n$ denotes the dimension of the flat spactime. 

\noindent The generator $\stackrel{(0)}{W}$, also called ``solution of the master equation'', is then introduced: 
\begin{eqnarray}
\stackrel{(0)}{W}\;\,=\int\left[{\cal{L}}_F + s_a\,\sum_a\phi^{*\m_1...\m_{s_a}}_a\partial_{\m_1}C^a_{\m_2...\m_{s_a}}\right]d^n x\quad.
\end{eqnarray} 
Let us define the antibracket (we denote collectively the fields and ghosts as $\Phi^i$ and the antifields as $\Phi^*_i$):
\begin{eqnarray}
(A,B)=\frac{\d^L A}{\d\Phi^i}\frac{\d^R B}{\d\Phi^*_i} - 
\frac{\d^L A}{\d\Phi^*_i}\frac{\d^R B}{\d\Phi^i}\quad.
\end{eqnarray} 
The generator satisfies: $\Big(\stackrel{(0)}{W},A\Big) = \bs A$ where $\bs=\d+\g$ is the BRST differential of the theory.

\subsection{Cohomology of $\g$}

In this section, we introduce our notation for the cohomology of $\g$ (whose elements are called \textit{invariants}). It has been showed \cite{Bekaert:2005ka} that the local functions of $H^*(\g)$ for a spin-$s_a$ Fronsdal theory in flat spacetime only depend on the antifields, the Fronsdal tensor $F^a_{\m_1...\m_{s_a}}$, the curvature tensor $K^a_{\m_1\n_1|...|\m_{s_a}\n_{s_a}}$ (which consists of $s_a$  
curls of the field) and their derivatives, as well as some non $\g$-exact ghost tensors, denoted $U^{(i)a}_{\m_1\n_1|...|\m_i\n_i|\n_{i+1}...\n_{s_a-1}}$ ($i<s_a$), that are the traceless part of the $i$ times antisymmetrised $i$th derivatives of the ghosts. For example ($i=1$): \begin{eqnarray}U^{(1)a}_{\m_1\n_1|\n_2...\n_{s_a-1}}=\partial_{[\m_1}C^a_{\n_1]\n_2...\n_{s_a-1}}-\frac{(s-2)}{n+2s-4}\eta^{}_{[\m_1|(\n_2}\partial^\r C^a_{\n_3...\n_{s_a-1})|\n_1]\r}\quad.\end{eqnarray} 
Of course, the zeroth tensor ($i=0$) is the undifferentiated ghost itself. 
More generally, the ghost tensors $U^{(i)a}_{\m_1\n_1|...|\m_i\n_i|\n_{i+1}...\n_{s_a-1}}$ give irreducible representations of the Lorentz algebra $\mathfrak{o}(n-1,1)$ labeled by Young diagrams made of two rows of respective lengths $s_a-1$ and $i<s_a$. This property will be extremely useful in order to classify the nonabelian cubic deformations.

In the case of a sum of Fronsdal theories, the cohomology of $\g$ is simply the direct product of the cohomologies of the separate theories. The cohomology is thus the following set of functions: \begin{eqnarray}H^*(\g)=\left\{f([\Phi^*_a],[F^a],[K^a],C^a,U^{(i)a})\right\}\quad,\end{eqnarray} where the square brackets around a field here denotes the corresponding field and all its derivatives. The functions $f$ are polynomials in the local case (and the number of derivatives is bounded from above).

\subsection{Consistent deformations}

The problem of consistently deforming a free theory, like Fronsdal's theory, into a full, interacting theory, can be reformulated within the antifield formalism \cite{Barnich:1993vg}: if one considers an expansion of the generator $W$ in terms of a parameter $g$: 
$W = \int w \,= \;\stackrel{(0)}{W}+\,g\stackrel{(1)}{W}+\,g^2\stackrel{(2)}{W}+\,...\,$, then the deformation is consistent if the full generator satisfies the {\textit{master equation}} $(W,W)=0$ to all orders in $g$. Since Fronsdal's theory is consistent, the initial generator satisfies $(\stackrel{(0)}{W},\stackrel{(0)}{W})=0$, which implies that $\bs$ is a differential ($\bs^2=0$). 

The first order equation is $(\stackrel{(0)}{W},\stackrel{(1)}{W})=\bs\stackrel{(1)}{W}\,=0\,$. In the case of a local deformation, $\stackrel{(1)}{W}$ must be the integral of a local $n$-form $a:=\,\stackrel{(1)}{w}$, and the equation 
$\bs\stackrel{(1)}{W}\,=0$ becomes a $\bs$-cocycle relation modulo ${\rm d}$: 
\begin{eqnarray}\bs \,a\,+\,{\rm d} \,b=0\quad,
\end{eqnarray}
where the operator ${\rm d}$ denotes the total exterior differential. 
Since $\bs$-exact and ${\rm d}$-exact terms in the cocycle $\stackrel{(1)}{w}$ correspond to trivial deformations, the first order inequivalent deformations are described by the cohomology class $H^{0,n}(\bs|{\rm d})\,$, 
see \cite{Barnich:1993vg,Henneaux:1998i}.

The obstructions under consideration in this paper arise when checking whether the first order vertices satisfy the master equation at second order, the local form of which reads:
\begin{eqnarray}(\stackrel{(1)}{w},\stackrel{(1)}{w})\,d^n x=-\frac{1}{2}\,\bs\stackrel{(2)}{w}+\,{\rm d}\,e
\label{soe}\quad.
\end{eqnarray}

\subsection{Cubic vertices}

It has been showed, for values of the spin up to 4, that the only first order nonabelian solutions of the local master equation in flat spacetime are cubic in the fields, ghosts or antifields \cite{Barnich:1994,Boulanger:2000rq,Bekaert:2006jf,Boulanger:2008tg}. For the spin-2 fields, this allows one to show \cite{Boulanger:2000rq} that Einstein's gravity is the only nonabelian consistent deformation with at most two derivatives of 
the free Pauli-Fierz theory in dimension $n>3\,$. For the spin-3 case, we 
know \cite{Bekaert:2006jf} that the BBvD vertex is one of the only two possible nonabelian first order deformations of the spin-3 Fronsdal theory. 
For values of the spin strictly greater than 4, it is still not proved 
within the antifield formalism whether other kinds of deformations are possible, for example starting with a quartic first order vertex. However, the problem consisting in computing cubic first order deformations can be addressed. As will be shown in the next subsection, the classification of the candidates is severe enough to put strong constraints on the allowed number of derivatives and on the gauge structure of the deformation, depending on the spins involved. Let us also mention the very powerful light-cone gauge method 
used in~\cite{Metsaev:2005ar,Metsaev:2007rn}. 

The components of the generator $w$ carry a ghost number $0$, which means that their antifield number is equal to their pure ghost number. 
Hence, for cubic deformations, only expressions with at most antifield number 2 will appear, since the pure ghost number of any individual field is at most 1
in Fronsdal's theory.
Therefore the cubic first order deformation can be 
expanded in the antifield number $a=a_0+a_1+a_2$ and the master equation decomposes into the 
following system of equations: 
\begin{eqnarray}
&\g \,a_2=0\label{eqagh2}& \\
&\d \,a_2+\g \,a_1+{\rm d }\, b_1=0\label{eqagh1}&\\
&\d \,a_1+\g \,a_0+{\rm d }\, b_0=0\label{eqagh0}&\quad.
\end{eqnarray}
The component $a_2$ contains the information about the first order deformation of the 
({\textit{a priori}} on-shell) gauge algebra. 
The nonabelian deformations are thus characterized by a non-vanishing $a_2$ component. 
A cubic $a_2$ is linear in the antifield number $2$ antifields and quadratic in the ghosts, it does not depend on the fields.
Consequently, the gauge algebra closes off-shell at first-order in the deformation, for cubic vertices. Moreover, the top form $a_2$
is $\g$-closed by Eq.(\ref{eqagh2}) and any $\g$-exact term is trivial in the sense that it is the antifield number $2$ part of an $\bs$-exact term in $a$. Thus $a_2$ can efficiently be written as a representative of $H^2(\g)$. Finally, $a_2$ is defined modulo ${\rm d }\,$, which allows one to only consider undifferentiated antifields. Therefore, without loss of generality, the general structure of $a_2$ reads schematically:
\begin{eqnarray}
a_2=f^a_{bc|(i)(j)}\,C^{*}_a\, U^{(i)b}\, U^{(j)c}\,d^n x\quad,
\end{eqnarray}
where $f^a_{bc|(i)(j)}$ are internal coefficients. This expression of $a_2$ encodes the structure constants of the gauge algebra, at first order in the  deformation. A Poincar\'e invariant $a_2$ is Lorentz-invariant (the spacetime indices must all be contracted) and does not explicitly depend  on the spacetime coordinates, therefore the coefficients $f^a_{bc|(i)(j)}$ are constants.

Finally, the component of maximal antifield number of the second order equation (\ref{soe}), 
which is the test that we use to exhibit the obstructions, reads: 
\begin{eqnarray}(a_2,a_2)\,d^n x = \g \,c_2+{\rm d }\, e_2\quad.
\label{secondorder}
\end{eqnarray}
This equation is the translation, within the antifield formalism, of the 
lowest-order component of the Jacobi identity for the gauge algebra. 

%******************************************************************************
\section{General results on the gauge algebra deformations \label{sec:genr}}
%*******************************************************************************

In this section, we provide general arguments that simplify the classification 
of the cubic nonabelian deformations for an arbitrary spin configuration. For 
any cubic configuration of the type $s-s'-s''$ (\textit{i.e.} including fields 
of respective spins $s$, $s'$ and $s''$), with $s\leqslant s'\leqslant s''$, 
there is a small number of possibilities of building consistent $a_2$ 
expressions, and only some of them are related to a consistent $a_1\,$. 
As we said previously, a cubic $a_2$ can always be written in the form 
$$a_2=C^* U^{(i)}\, U^{(j)}d^n x+\g(...)\quad,$$ 
where the $U^{(i)}$ are non-$\g$-exact ghost tensors. 
A strong constraint on such candidates $a_2$ is that the product 
$U^{(i)} U^{(j)}$ of ghost tensors (in general, with implicit contractions of 
indices) must be contracted with the antighost $C^*$ which is itself a 
symmetric Lorentz tensor. The Littlewood-Richardson rules will be used 
throughout this section in order to analyze all possible contractions of the 
indices from the two ghost tensors and the antifield.\footnote{More precisely, 
the specific rules for product and division of Young diagrams are applied here 
(see \textit{e.g.} the appendix A of \cite{Bekaert:2005ka} for a self-contained 
review of these Littlewood-Richardson rules).} With the help of these rules, we 
will show that the previous constraint implies several strong conditions on the 
allowed values of the numbers of derivatives $i$, $j$ and of spins $s$, $s'$ 
and $s''\,$.

\subsection{Product of ghost tensors}\label{product}

Let us consider a product of two ghost tensors $U^{(i)}$ and $U^{(j)}$, corresponding respectively  to spin $s_1$ and $s_2$ (with $s_1\leqslant s_2$). 

\begin{enumerate}
\item[\textbf{A.}] Firstly, we may study the minimal number of free indices in that product (in other words, the maximal number of contracted indices).    
	  \begin{itemize}
	  \item[\textbf{A.1.}] If $i\leqslant j$, all of the indices of 
	  $U^{(i)}$ can be contracted with $s_1+i-1$ indices of $U^{(j)}$. Let us 
	  visualize in terms of Young diagrams the symmetry properties of the tensors 
	  resulting from the maximal contraction of indices:
\begin{eqnarray}U^{(i)}: \begin{picture}(60,0)(0,0)\multiframe(0,0)(10,0){1}(50,10){$s_1-1$}\multiframe(0,-10.5)(10,0){1}(20,10){$i$}\end{picture},\ U^{(j)}:\begin{picture}(70,0)(0,0)\multiframe(0,0)(10,0){1}(60,10){$s_2-1$}\multiframe(0,-10.5)(10,0){1}(40,10){$j$}\end{picture}\nonumber\end{eqnarray}\begin{eqnarray} &\Rightarrow  {\textrm{Maximal contraction}}:&\bullet\ \begin{picture}(36,0)(0,0)\multiframe(0,-2)(10,0){1}(35,10){$s_2-s_1$}\end{picture}\otimes\begin{picture}(30,0)(0,0)\multiframe(0,-2)(10,0){1}(30,10){$j-i$}\end{picture}\textrm{ if $j<s_1$}\nonumber\\&&\bullet\ \bigoplus_a\begin{picture}(105,0)(0,0)\multiframe(0,1)(10,0){1}(100,10){$s_2-s_1+j-i-a$}\multiframe(0,-9.5)(10,0){1}(20,10){$a$}\end{picture}\textrm{ if $j-i\geqslant a\geqslant j-s_1+1>0$ }\nonumber\quad.\end{eqnarray}

    Since $U^{(j)}$ bears $s_2+j-1$ indices, the minimal number $N_{min}$ of 
    free indices is $s_2-s_1+j-i$. 
    Furthermore, these free indices can be symmetrized if $j<s_1$ since there 
    is a component \begin{picture}(85,0)(0,0)\multiframe(0,-2)(10,0){1}(80,10){$s_2-s_1+j-i$}\end{picture} 
    in the tensor product. If $j\geqslant s_1$, no contraction of the two 
    tensors $U$ can be symmetrized and thus no Lorentz invariant $a_2$ can be 
    built. Consequently, 
    $$\mbox{max}\{i,j\}<s_1$$ in order to have symmetrizable free indices, as 
    can be seen for the other case as well.

    \item[\textbf{A.2.}] 
    If $j<i<s_1\leqslant s_2$, let us visualize the ghost tensors: 
\begin{eqnarray}U^{(i)}: \begin{picture}(80,0)(0,0)\multiframe(0,0)(10,0){1}(75,10){$s_1-1$}\multiframe(0,-10.5)(10,0){1}(20,10){$j$}\multiframe(20.5,-10.5)(10,0){1}(40,10){$i-j$}\end{picture},\ U^{(j)}:\begin{picture}(120,0)(0,0)\multiframe(0,0)(10,0){1}(75,10){$s_1-1$}\multiframe(75.5,0)(10,0){1}(40,10){$s_2-s_1$}\multiframe(0,-10.5)(10,0){1}(20,10){$j$}\end{picture}\nonumber\quad.\end{eqnarray}
    The maximal contraction is obtained by contracting the $s_1-1$ boxes and 
    the $j$ boxes, which leaves one with a product: \begin{picture}(36,0)(0,0)\multiframe(0,-2)(10,0){1}(35,10){$s_2-s_1$}\end{picture} $\otimes$ \begin{picture}(30,0)(0,0)\multiframe(0,-2)(10,0){1}(30,10){$i-j$}
    \end{picture} , 
    which always involves a totally symmetric component. Explicitly, this reads:
\begin{eqnarray}
U^{(i)\mu_1\nu_1|...|\mu_j\nu_j|\mu_{j+1}\b_1|...|\mu_i\b_{i-j}|\mu_{i+1}...\mu_{s_1-1}}U^{(j)}_{\mu_1\nu_1|...|\mu_j\nu_j|\mu_{j+1}...\mu_{s_2-1}}\quad.
\end{eqnarray} 
The $\b$ indices are free and there are $s_2-s_1$ free $\mu$ indices. The minimal number $N_{min}$ of free indices in this case is thus $s_2-s_1+i-j$.\end{itemize} The two cases can be gathered as \begin{eqnarray}N_{min}=s_2-s_1+|i-j|.\end{eqnarray}

\item[\textbf{B.}] 
Secondly, the maximal number of free indices that can be symmetrized in a 
product $U^{(i)}U^{(j)}$ may also be studied: 
   \begin{itemize} 
   \item[\textbf{B.1.}] If $i\leqslant j<s_1$, then $j$ pairs have to be contracted:\begin{eqnarray}U^{(i)}:\begin{picture}(95,0)(0,0)\multiframe(0,0)(10,0){1}(30,10){$j$}\multiframe(30.5,0)(10,0){1}(60,10){$s_1-j-1$}\multiframe(0,-10.5)(10,0){1}(20,10){$i$}\end{picture},\ U^{(j)}:\begin{picture}(115,0)(0,0)\multiframe(0,0)(10,0){1}(110,10){$s_2-1$}\multiframe(0,-10.5)(10,0){1}(30,10){$j$}\end{picture}\quad.\end{eqnarray}
   If one contracts less than $j$ pairs of indices, 
   some indices remain in the second line of $U^{(j)}$ and the result cannot 
   contain any totally symmetric component. This leaves us with $s_1+s_2+i-j-2$ 
   free indices. 
   \item[\textbf{B.2.}] If $i\geqslant j$, in the same way $i$ pairs have to be contracted, leaving $s_1+s_2+j-i-2$ free indices.\end{itemize} 
   Thus, the maximal number of free and symmetrizable indices in the tensor 
   product of the two ghost tensors is 
   \begin{eqnarray}N_{max}=s_1+s_2-|i-j|-2.
   \end{eqnarray}
\end{enumerate}

\subsection{Bound on the difference of the number of derivatives}

After these general considerations on allowed products of two ghost tensors $U^{(i)}$ and $U^{(j)}$, let us now consider again a candidate for $a_2$, for a configuration $s-s'-s''$ with $s\leqslant s'\leqslant s''$. 

Firstly, it is noticed that there are no nonabelian deformations if $s''\geqslant s+s'$. For example, there is no way of building a $1-1-s$ deformation if $s\geqslant 2$, or a $2-2-s$ deformation with $s\geqslant 4$. This property comes from\footnote{Notice that this property follows from purely diagrammatic reasoning and therefore also applies in $AdS$. Indeed, it can be checked that the structure constants of the higher-spin algebra in \cite{Vasiliev:2003ev} obey to this bound.} the fact that the product of two Young diagrams whose first rows have lengths $s$ and $s'$ cannot contain a Young diagram whose first row has length 
length $s''\geqslant s+s'\,$.

Secondly, we can show a stronger property that involves the numbers of derivatives $i$ and $j$. Three cases have to be studied, related to the spin of the antifield. In the case of a spin-$s$ antifield $C^{*\mu_1...\mu_{s-1}}$, the minimum number of free indices in the product $U^{(i)}U^{(j)}$ is $s''-s'+|i-j|$. In order for $a_2$ to be Lorentz-invariant, every index must be contracted, hence the former number must be lower or equal to the number of indices of the antifield. We thus obtain the relation: $s''-s'+|i-j|\leqslant s-1$. In the case of a spin $s'$ antifield, the same argument can be applied, it leads to the relation $s''-s+|i-j|\leqslant s'-1$, which is the same as the first one. Finally, in the case of a spin $s''$ antifield, the minimal condition $s'-s+|i-j|\leqslant s''-1$ is always satisfied, since $i<s$ and $j<s$ imply $|i-j|-s<0$ while, moreover, $s'\leqslant s''$.
On the other hand, in this case, we have to consider the fact that the maximum number of free symmetrizable indices must be greater or equal than the number of indices of the antifield: $s+s'-|i-j|-2\geqslant s''-1$, and we obtain once again the same condition. 

Thus for any combination of the fields, the spins have to satisfy the inequality: \begin{eqnarray}s+s'-s''>|i-j|\geqslant0\quad.\end{eqnarray} 
This provides an upper bound on the difference between the numbers of derivatives in the two ghost tensors.

\subsection{Conditions on the total number of derivatives}

If we want to build Lorentz-invariant and parity-even expressions, the total number of indices has to be even. For an antifield of spin $s_3$ and ghost tensors $U^{(i)}$ of spin $s_1$ and $U^{(j)}$ of spin $s_2$, the numbers of indices are $s_3-1$, $s_1+i-1$ and $s_2+j-1$, for a total of $s_1+s_2+s_3+i+j-3$. Thus, we find that, for a configuration $s\leqslant s'\leqslant s''\,$:
\begin{eqnarray}
s+s'+s''+i+j\equiv 1(mod\ 2)\quad.
\end{eqnarray}

Furthermore, let us emphasize that the total number of derivatives $\,i+j\,$ 
is bounded. As was mentioned in Subsection \ref{product}, the numbers $i$ and $j$ must be strictly lower than the spins of the two ghost tensors in order for the free indices to be symmetrizable. If $U^{(i)}$ is of spin $s_1$ and $U^{(j)}$ is of spin $s_2$ with $s_1\leqslant s_2$, then $i\leqslant s_1-1$ and $j\leqslant s_1-1$. Thus, we obtain the condition $i+j\leqslant 2s_1-2\,$. 
If we consider a candidate for $a_2\,$, this condition immediately tells that 
\begin{eqnarray}
i+j\leqslant 2s'-2\quad.
\end{eqnarray} 
More, precisely, if the spin-$s$ antifield is considered, 
then the upper bound is $2s'-2\,$. 
If either the spin $s'$ or $s''$ antifield is considered, the upper bound 
is even lower: $i+j\leqslant 2s-2\,$. 

\subsection{A general theorem and a particular candidate}\label{thma2}

Let us summarize all previous considerations in the following theorem:
\newpage 

\begin{theorem}Given a cubic configuration of fields with spins $s\leqslant s'\leqslant s''$, the possible Poincar\'e invariants $a_2=C^* U^{(i)}U^{(j)}\,d^nx$ are contractions of an undifferentiated antifield number-$2$ antighost and of two ghost tensors, involving $i$ and $j$ derivatives. The spins and the numbers of derivatives have to satisfy the following properties:\begin{itemize}
\item $0\leqslant |i-j|<s+s'-s''$
\item $s+s'+s''+i+j$ is odd
\item In the case of a spin-$s$ antifield: $i+j\leqslant 2s'-2$\\ In the case of a spin $s'$ or $s''$ antifield: $i+j\leqslant 2s-2$
\end{itemize}
\end{theorem}

To end up this section, let us mention that the candidate $a_2$ with the highest number of derivatives $i+j=2s'-2$ always satisfies 
Eq. (\ref{secondorder}) due to the large number of derivatives
involved in it. 
Let us also show that the same candidate $a_2$ satisfies 
Eq.(\ref{eqagh1}), \textit{i.e.} that a corresponding first-order deformation of the gauge transformations exist. This is less obviously seen than the previous property: We do not provide the corresponding $a_1$ explicitly but the equation ensures that it exists. In the case of an even number of derivatives (in other words when the sum $s+s'+s''$ is odd), the candidate with $2s'-2$ derivatives reads: 
\begin{eqnarray}a_2&=&C^{*\mu_1...\mu_{s-1}}U^{(s'-1)}_{\a_1\r_1|...|\a_\l\r_\l|\mu_1\r_{\l+1}|...|\m_{s'-\l-1}\r_{s'-1}}\times\nonumber\\&&\times\, U^{(s'-1)\a_1\r_1|...|\a_\l\r_\l|\phantom{\m_{s'-\l}}\r_{\l+1}|...|\phantom{\mu_{2s'-2\l-2}}\r_{s'-1}|}_{\,\phantom{(s'-1)\a_1\r_1|...|\a_\l\r_\l|}\m_{s'-\l}\phantom{\r_{\l+1}|...|}\m_{2s'-2\l-2}\phantom{\r_{s'-1}|}\m_{2s'-2\l-1}...\m_{s-1}}\,d^n x\quad,\end{eqnarray}
where $\l=\frac{s'+s''-s-1}{2}$. In terms of Young diagrams, this contraction can be seen as follows: \begin{eqnarray}
C^*:\begin{picture}(125,0)(0,0)\multiframe(0,0)(10,0){1}(50,10){$s'-\l-1$}\multiframe(50.5,0)(10,0){1}(70,10){$s''-\l-1$}\end{picture} , U^{(s'-1)}_{s'}:\begin{picture}(75,0)(0,0)\multiframe(0,0)(10,0){1}(20,10){$\l$}\multiframe(20.5,0)(10,0){1}(49.5,10){$s'-\l-1$}\multiframe(0,-10.5)(10,0){1}(70,10){$s'-1$}\end{picture} , U^{(s'-1)}_{s''}:\begin{picture}(85,0)(0,0)\multiframe(0,0)(10,0){1}(20,10){$\l$}\multiframe(20.5,0)(10,0){1}(70,10){$s''-\l-1$}\multiframe(0,-10.5)(10,0){1}(70,10){$s'-1$}\end{picture}\quad.\end{eqnarray} 
The variation of this expression under delta takes the form: 
\begin{eqnarray}
\nonumber
\d a_2=d(...)+s\Big[\phi^{*\mu_1...\mu_s}
-\frac{(s-1)(s-2)}{2(n+2s-6)}\eta^{(\m_1\m_2}
\phi^*{}'{}^{\m_3...\m_{s-1})\m_s}\Big]\partial_{\m_s}
\Big[U^{(s'-1)}U^{(s'-1)}\Big]\quad.
\end{eqnarray}
The action of $\partial_{\m_s}$ on the spin $s'$ tensor $U^{(s'-1)}$ is automatically $\g$-exact because it is not possible to take one more curl. Actually, the action of $\partial_{\m_s}$ on the spin $s''$ tensor $U^{(s'-1)}$ is also $\g$-exact because the contraction of all free $\mu$ indices with the symmetric indices of the factor linear in the antifield $\phi^*$ prevent any more curl.
 
The case of an even sum $s+s'+s''$ is a bit more complicated. There are two possible terms, that have to be proportional in order for $a_1$ to exist: \begin{eqnarray}a_2&=&\alpha\, C^{\mu_1...\mu_s}U^{(s'-1)}_{\a_1\r_1|...|\a_\l\r_\l|\mu_1\r_{\l+1}|...|\m_{s'-\l-1}\r_{s'-1}}\times\nonumber\\&&\quad\times\, U^{(s'-2)\a_1\r_1|...|\a_\l\r_\l|\phantom{\m_{s'-\l}}\r_{\l+1}|...|\phantom{\mu_{2s'-2\l-3}}\r_{s'-2}|\r_{s'-1}}_{\,\phantom{(s'-2)\a_1\r_1|...|\a_\l\r_\l|}\m_{s'-\l}\phantom{\r_{\l+1}|...|}\m_{2s'-2\l-3}\phantom{\r_{s'-2}|\r_{s'-1}}\m_{2s'-2\l-2}...\m_{s-1}}\,d^n x\ \nonumber\\&&
+\,\beta\, C^{\mu_1...\mu_s}U^{(s'-2)}_{\a_1\r_1|...|\a_\l\r_\l|\mu_1\r_{\l+1}|...|\m_{s'-\l-2}\r_{s'-2}|\r_{s'-1}}\times\nonumber\\&&\ \,\quad\times\, U^{(s'-1)\a_1\r_1|...|\a_\l\r_\l|\phantom{\m_{s'-\l}-1}\r_{\l+1}|...|\phantom{\mu_{2s'-2\l-2}}\r_{s'-1}|}_{\,\phantom{(s'-1)\a_1\r_1|...|\a_\l\r_\l|}\m_{s'-\l-1}\phantom{\r_{\l+1}|...|}\m_{2s'-2\l-3}\phantom{\r_{s'-1}|}\m_{2s'-2\l-2}...\m_{s-1}}\,d^n x\quad,\end{eqnarray} where $\l=(s'+s''-s-2)/2$
and $\alpha,\beta$ are coefficients. In terms of Young diagrams, these contractions read:\begin{eqnarray}&&
C^*:\begin{picture}(125,0)(0,0)\multiframe(0,0)(10,0){1}(50,10){$s'-\l-1$}\multiframe(50.5,0)(10,0){1}(70,10){$s''-\l-2$}\end{picture} , U^{(s'-1)}_{s'}:\begin{picture}(75,0)(0,0)\multiframe(0,0)(10,0){1}(20,10){$\l$}\multiframe(20.5,0)(10,0){1}(50,10){$s'-\l-1$}\multiframe(0,-10.5)(10,0){1}(60,10){$s'-2$}\multiframe(60.5,-10.5)(10,0){1}(10,10){1}\end{picture} , U^{(s'-2)}_{s''}:\begin{picture}(100,0)(0,0)\multiframe(0,0)(10,0){1}(20,10){$\l$}\multiframe(20.5,0)(10,0){1}(70,10){$s''-\l-2$}\multiframe(91,0)(10,0){1}(10,10){1}\multiframe(0,-10.5)(10,0){1}(60,10){$s'-2$}\end{picture}\nonumber\quad,\\ \textrm{and}&&\nonumber\\&& C^*: \begin{picture}(125,0)(0,0)\multiframe(0,0)(10,0){1}(50,10){$s'-\l-2$}\multiframe(50.5,0)(10,0){1}(70,10){$s''-\l-1$}\end{picture} , U^{(s'-2)}_{s'}:\begin{picture}(85,0)(0,0)\multiframe(0,-10.5)(10,0){1}(20,10){$\l$}\multiframe(20.5,-10.5)(10,0){1}(49.5,10){$s'-\l-2$}\multiframe(0,0)(10,0){1}(80,10){$s'-1$}\end{picture} , U^{(s'-1)}_{s''}:\begin{picture}(90,0)(0,0)\multiframe(0,0)(10,0){1}(20,10){$\l$}\multiframe(20.5,0)(10,0){1}(70,10){$s''-\l-1$}\multiframe(0,-10.5)(10,0){1}(80,10){$s'-1$}\end{picture}\quad.\nonumber
\end{eqnarray}
This time, the computation of $\d a_2$ consists of four terms. The term involving the derivative of the spin-$s'$ tensor $U_{s'}^{(s'-1)}$ is automatically $\g$-exact, and the term where the spin-$s''$ tensor $U_{s''}^{(s'-1)}$ is differentiated is $\g$-exact, thanks to the same arguments as for the odd case. On the other hand, the terms where the $U^{(s'-2)}$ tensors are differentiated are problematic. Fortunately, the non-$\g$-exact terms that appear are the same in the two expressions and the coefficients $\a$ and $\b$ can be fitted to obtain a $\g$-exact result.

%***********************************************************************************
\section{Proof of the strong obstruction on the BBvD vertex}\label{sec:obstr}
%***********************************************************************************

The BBvD first order deformation \cite{Berends:1984wp} has been obtained in the antifield formulation in \cite{Bekaert:2006jf}. We denote the spin-3 ghost tensors $T^A_{\m\n|\r}:=U^{(1)A}_{\m\n|\r}$ and $U^A_{\m\n|\r\s}:=U^{(2)A}_{\m\n|\r\s}$. The antifield number $\ 2$ component $a_2$ of the BBvD deformation contains $2$ derivatives and reads:
\begin{eqnarray}a_{2,BBvD}=f^A_{\phantom{A}BC}C^{*\m\n}_A\left[T^B_{\mu\a|\b}T^{C\,\a|\b}_\n-2\,T^B_{\mu\a|\b}T^{C\,\b|\a}_\n+\frac{3}{2}\,C^{B\a\b}U^C_{\m\a|\n\b}\right]d^n x+\g c_2\quad,\end{eqnarray} where the capital internal indices span the multiplet of spin-3 fields.
The corresponding cubic vertex $a_{0,BBvD}$ contains $3$ derivatives.
It has been showed \cite{Bekaert:2006jf} that the second order expression, $(a_{2,BBvD},a_{2,BBvD})$ as in Equ.(\ref{secondorder}), presents an obstruction containing terms of the structure $C^*TTU$ and $C^*CUU$, that are not $\g$-exact modulo $d$, and cannot be eliminated. The coefficient of the obstruction is $f_{ABC} f^A_{DE}$, whose vanishing implies the vanishing of the deformation itself. Let us notice that, in dimension 3, since $U^A_{\a\b|\c\d}\equiv0$, the BBvD candidate passes the test. In dimension 4, some Schouten identities could imply the weaker associativity condition $f^{}_{AB[C}f^A_{D]E}=0$, however, this still implies the vanishing of $f_{ABC}$ in the end. 

Furthermore, a new nonabelian cubic vertex with 5 derivatives was found in \cite{Bekaert:2006jf}. However, it vanishes in dimension 3 where the BBvD candidate therefore involves the maximal possible number of derivatives. In dimension 4, it has been showed that Schouten identities imply the vanishing of the corresponding component of $a_2$, thus the 5-derivative deformation is Abelian in that case. 
Let us remark that the case of dimension 3 is a bit special, since traceless tensors associated with a Young diagram whose first two columns have length 2 identically vanish.\footnote{More generally, in dimension $n$, any tensor associated with an irreducible representation of $\mathfrak{o}(n)$ (and thus traceless), whose Young diagram is such that the sum of the heights of the first two columns is greater than $n$, identically vanish (see \cite{Hamermesch}, page 394). Let us remark that, for $n\geqslant 4$, two-row tensors, such as the traceless part of the curvature or the strictly non-$\g$-exact ghost tensors, are never constrained by this condition.} This implies that every fieldstrength vanishes on-shell, which only allows topological theories, for spin $\geqslant 2\,$.

We now want to prove that no other $a_2$'s can provide the same kind of terms that could compensate the obstruction. First, the antibracket $(a_{2,BBvD},a_{2,BBvD})$ is of course quartic in the spin-3 fields (in the extended meaning of fields, ghosts or antifields) and it contains four derivatives. It can only be compensated by terms in another antibracket $(a_2,a_2)$ which have exactly the same structure. The only possibility of getting terms quartic in the spin-3 fields is to take the $(a_2,a_2)$ of two expressions with the same spin configuration $s-3-3$. Then, the first rule of our general theorem ensures that $1\leqslant s \leqslant 5$. The nonabelian $1-3-3$ and $2-3-3$ deformations have been completely classified \cite{Boulanger:2006gr,Boulanger:2008tg}: in both cases, there is only one solution, whose $a_2$ is linear in the antifield with lowest spin (1 or 2) and antifield number $2$. These candidates satisfy trivially $(a_2,a_2)=0$, hence they cannot help for the BBvD obstruction.

To complete the argument, we have to investigate the $3-3-4$ and $3-3-5$ cases. It is rather simple: we will prove that the only $a_2$ candidates that are related to an $a_1$ contain 
at least three derivatives, which is sufficient to be sure that no obstruction containing four derivatives will arise. The results about those two cases are presented in the next two subsections. The results are sufficient to establish the inconsistency of the BBvD deformation in dimension greater than three, in the parity-invariant case, thereby invalidating the hopes expressed by the authors of \cite{Berends:1984wp,Berends:1984rq}
concerning a possible solution of their problem by the addition of 
totally symmetric higher spin contributions. Therefore, in flat spacetime, their spin-3 self coupling is definitely inconsistent and no 
totally symmetric higher spin field can cure this problem contrary to the general belief. It is only in $(A)dS$ that this candidate can play a role, as suggested by the Fradkin-Vasiliev cubic vertices \cite{Fradkin:1987ks,Fradkin:1987qy}.

\vspace{1mm}
\noindent{\bf{Remark:}} The $a_2$ components that are considered in the sequel for the $3-3-4$ and $3-3-5$ cases are not proved to be part of consistent first order solutions. Anyway, since we seek a negative result, it is obvious that, if the obstruction remains when considering all of the candidates, it would remain \textit{a fortiori} if these candidates are obstructed at first order.

\subsection{Study of $a_2$ in the $3-3-4$ case}

Let us use the theorem of Subsection \ref{thma2}, with $s=s'=3$ and $s''=4$. The sum of the spins is even, thus the number of derivatives in $a_2$ has to be odd. The maximum is $2s'-3=3$. Furthermore, the difference between the numbers of derivatives acting on the two ghosts obeys $|i-j|<s+s'-s''=2$, and is thus equal to 1. The possible strictly non-$\g$-exact Lorentz-invariant expressions with one derivative read: \begin{eqnarray}\stackrel{(1)}{t}{}^{\!\!AB}=C^{*\m\n\r}C^{A\a}_{\m}T^B_{\a\n|\r}\ ,\ \stackrel{(2)}{t}{}^{\!\!AB}=C^{*A\m\n} T^B_{\m\a|\b} C_\n^{\ \a\b}\ ,\ \stackrel{(3)}{t}{}^{\!\!AB}=C^{*A\m\n} C^{B\r\s} U^{(1)}_{\m\r|\n\s}\quad.\end{eqnarray} Those with three derivatives are: \begin{eqnarray}\stackrel{(4)}{t}{}^{\!\!AB}=C^{*\m\n\r}T_{\m}^{A\a|\b}U_{\n\a|\r\b}^B\ ,\ \stackrel{(5)}{t}{}^{\!\!AB}=C^{*A\m\n}T^{B\a\b|\c}U^{(2)}_{\a\b|\c\m|\n}\ ,\ \stackrel{(6)}{t}{}^{\!\!AB}=C^{*A\m\n}U^B_{\a\b|\c\m}U^{(1)\a\b|\c}_{\phantom{(1)\a\b|\c}\n}\quad.\end{eqnarray}The spin-4 internal indices have not been written explicitly since no symmetries can arise involving them (similarly, in the next section, the spin-5 indices are not written as well).
Let us check that the candidates with three derivatives are related to an $a_1$: \begin{eqnarray}\d\stackrel{(4)}{t}{}^{\!\!AB}=\mbox{divergence}+\g(...)+4\phi^{*\m\n\r\s}U^{A\a|\phantom{\s}\b}_{\m\phantom{\a|}\s}U^B_{\n\a|\r\b}-\frac{6}{n+2}\phi^*{}'{}^{\r\s}U^{A\n\a|\phantom{\s}\b}_{\phantom{A\n\a|}\s}U^B_{\n\a|\r\b}\quad.\end{eqnarray} This term is antisymmetric in $AB$, thus a symmetric set of coefficients ensures the vanishing of the non-$\g$-exact terms. The variation under $\d$ of the two other terms provides the same non-$\g$-exact term $\phi^{*A\m\n\r}U^{B\a\b|\phantom{\r}\c}_{\phantom{B\a\b|}\r}U^{(2)}_{\a\b|\c\m|\n}$, so they vanish if $\stackrel{(5)}{t}{}^{\!\!AB}$ and $\stackrel{(6)}{t}{}^{\!\!AB}$ have opposite coefficients. Finally, we get as candidates with three derivatives: \begin{eqnarray}a_{2,3}=k_{(AB)}C^{*\m\n\r}T_{\m}^{A\a|\b}U_{\n\a|\r\b}^B d^n x+l_{AB}C^{*A\m\n}\Big[T^{B\a\b|\c}U^{(2)}_{\a\b|\c\m|\n}-U^B_{\a\b|\c\m}U^{(1)\a\b|\c}_{\phantom{(1)\a\b|\c}\n}\Big]d^n x\quad.\end{eqnarray} On the other hand, the candidates involving one derivative are obstructed: \begin{eqnarray}\d \stackrel{(1)}{t}{}^{\!\!AB}=\mbox{divergence}+\g(...)+4\Big(\phi^{*\m\n\r\s}-\frac{3}{n+2}\eta^{(\m\n}\phi^*{}'{}^{\r)\s}\Big)\Big[-T^{A\a}_{\phantom{A\a}\s|\m}T^B_{\a\n|\r}+C^{A\a}_{\m}U^B_{\a\n|\s\r}\Big]\quad.\end{eqnarray} All the terms vanish if $\stackrel{(1)}{t}{}^{\!\!AB}$ is multiplied by a symmetric coefficient, except one proportional to the trace of $\phi^*$: $\frac{-4}{n+2}\phi^*{}'{}^{\n\s}C^{A\a\r}U^{B}_{\a\n|\s\r}$. This obstruction cannot be removed. The variation under $\d$ of $\stackrel{(2)}{t}{}^{\!\!AB}$ and $\stackrel{(3)}{t}{}^{\!\!AB}$ contains the obstructions $\phi^{*A\m\n\r}U^B_{\m\a|\r\b}C_\n^{\phantom{\n}\a\b}$ and $\phi^{*A\m\n\r}C^{\a\b}U^{(2)}_{\m\a|\n\b|\r}$. Finally, the only possible $3-3-4$ deformation contains three derivatives in $a_2$. Even if the vertex exists, which is not sure, the only terms in $(a_{2,3},a_{2,3})$ contain six derivatives. This cannot remove the obstruction of the BBvD deformation.

\subsection{Study of $a_2$ in the $3-3-5$ case}

The theorem of Subsection \ref{thma2} ensures that the number of derivatives in $a_2$ is even, and is not greater than 4. Furthermore the two ghosts bear the same number of derivatives, since $|i-j|<3+3-5=1$. There are candidates with four, two and zero derivatives. Once again, only the candidates with four derivatives satisfy Eq.(\ref{eqagh1}). The possible terms with no derivatives are: \begin{eqnarray} \stackrel{(1)}{u}{}^{\!\!AB}=C^{*\m\n\r\s}C^A_{\m\n}C^B_{\r\s}\quad{\textrm{and}}\quad \stackrel{(2)}{u}{}^{\!\!AB}=C^{*A\m\n}C^{B\r\s}C_{\m\n\r\s}\quad.\end{eqnarray} 
Those with two derivatives are: \begin{eqnarray}\stackrel{(3)}{u}{}^{\!\!AB}=C^{*\m\n\r\s}T^{A\a}_{\phantom{A\a}\m|\n}T^B_{\a\r|\s}\quad{\textrm{and}}\quad \stackrel{(4)}{u}{}^{\!\!AB}=C^{*A\m\n}T^{B\a\b|\c}U^{(1)}_{\a\b|\c\m\n}\quad.\end{eqnarray} Those with four derivatives are: \begin{eqnarray} \stackrel{(5)}{u}{}^{\!\!AB}=C^{*\m\n\r\s}U^{A\a|\phantom{\n}\b}_{\m\phantom{\a|}\n}U^B_{\r\a|\s\b}\quad{\textrm{and}}\quad\stackrel{(6)}{u}{}^{\!\!AB}=C^{*\m\n}U^{\a\b|\c\d}U^{(2)}_{\a\b|\c\d|\m\n}\quad.\end{eqnarray}Let us notice that $\stackrel{(1)}{u}{}^{\!\!AB}$, $\stackrel{(3)}{u}{}^{\!\!AB}$ and $\stackrel{(5)}{u}{}^{\!\!AB}$ are naturally antisymmetric over $AB$. It is quite obvious that $\d\stackrel{(5)}{u}{}^{\!\!AB}$ is $\g$-exact modulo $d$ because the third derivative of the spin-3 ghost are $\g$-exact. Then, we can consider $\d\stackrel{(6)}{u}{}^{\!\!AB}$, which is $\g$-exact modulo $d$, for the same reason than the previous one and because $\partial_{(\r}U^{(2)\a\b|\c\d|}_{\phantom{(2)\a\b|\c\d|}\m\n)}$ is $\g$-exact. On the other hand, obstructions arise for any of the other candidates. For $\stackrel{(3)}{u}{}^{\!\!AB}$, one of the trace term remains, which is proportional to $\phi^*{}'{}^{\m\n\t}U^A_{\a\m|\t\s}T^{B\a\ |\s}_{\phantom{B\a}\n}$. For $\stackrel{(4)}{u}{}^{\!\!AB}$, the obstruction consists of two terms, proportional to $\phi^{*A\m\n\r}T^{B\a\b|\c}U^{(2)}_{\a\b|\c\m|\n\r}$ and $C^{*A\m\n\r}U^{B\a\b|\phantom{\r}\c}_{\phantom{B\a\b|}\r}U^{(1)}_{\a\b|\c\m\n}$. Finally, with no derivatives, the obstruction of $\stackrel{(1)}{u}{}^{\!\!AB}$ arises once again in the trace terms, it is proportional to $\phi^*{}'{}^{\m\n\r}T^A_{\a\m|\n}C^{B\a}_{\phantom{B\a}\r}$. The obstruction of $\stackrel{(2)}{u}{}^{\!\!AB}$ consists of two terms proportional to $\phi^{*A\m\n\r}C^{B\a\b}U^{(1)}_{\r\a|\b\m\n}$ and $\phi^{*A\m\n\r}T^{B\a|\b}_{\r}C_{\m\n\a\b}$. None of those obstructions can be removed, the only possible cubic $a_2$ thus involves four derivatives. Thus, any $(a_2,a_2)$ term involves eight derivatives, this can of course not remove the BBvD obstruction. Since we have considered a spin greater than four, we are not sure if the cubic deformations are the only possible ones, but any solution of degree higher than three will provide terms of power higher than four in $(a_2,a_2)$, which can not compensate the BBvD obstruction either.

\section{Conclusion}\label{concl}

Within the antifield formalism and in the case of cubic vertices between symmetric tensor gauge fields of any integer spins, we have introduced a set of criteria for the construction of consistent 
deformations of the gauge algebra.
Equivalently, these criteria are conditions on the structure constants of the 
gauge algebra at first order in the coupling constants. We have then showed 
that the Berends--Burgers--van Dam spin-3 vertex is obstructed at second order 
in the coupling constants, even if one introduces other symmetric tensor gauge 
fields in the theory. This invalidates, in Minkowski spacetime, the argument 
according to which the obstructions arising for a given set of values of the 
spins can be cured by terms involving higher values. This argument is related 
to the standard lore that an infinite tower of fields with unbounded spin is 
needed in any consistent higher-spin gauge theory. While this general 
expectation is not questioned, our result 
confirms some doubts about the mere existence of any consistent nonabelian 
Lagrangian formulation for higher-spin 
gauge fields in four-dimensional\footnote{The pure spin-$3$ non Abelian cubic 
vertex found in 
\cite{Bekaert:2006jf} 
exists only in higher dimensions ($n>4$) but this one is \textit{not} 
obstructed at the level of the gauge algebra. Nevertheless, the existence of a 
corresponding quartic vertex remains an open issue.} Minkowski spacetime, which 
would be obtained as a perturbative local deformation of Fronsdal's theory.
For flat spacetimes of higher dimensions, our results suggest that nonabelian cubic vertices containing only totally symmetric gauge fields and involving a number of derivatives which does \emph{not} saturate the upper bound that we found, would be inconsistent.
More precisely, our argument essentially
relies on the numbers of derivatives. Consistent first order vertices must involve a minimal number of derivatives. This minimum number increases with the values of the spin of the three fields contained in the cubic vertex. The number of derivatives is a good grading in flat spacetime, so the second-order equations involving different types of vertices are most of the times linearly independent (because they contain different numbers of derivatives). We can conjecture that many consistent cubic deformations in Minkowski spacetime are strongly obstructed in the same way.
However, in $(A)dS$ the number of derivatives is not a proper grading
and we expect that the obstructions exhibited for the flat-spacetime vertices 
do not show up in $(A)dS\,$, so that the Fradkin--Vasiliev cubic Lagrangian 
\cite{Fradkin:1987ks,Fradkin:1987qy} (see also \cite{Vasiliev:2001wa,Alkalaev:2002rq})
could be completed to give a fully consistent nonabelian Lagrangian theory to all orders in the coupling constant.

\section*{Acknowledgments}

We thank G. Barnich, M. Henneaux, A. Sagnotti, Ph. Spindel and P. Sundell 
for discussions. 
N.B. is a	Research Associate of the Fonds de la Recherche Scientifique--FNRS, Belgium. 

\providecommand{\href}[2]{#2}\begingroup\raggedright\endgroup

%\bibliographystyle{utphys}

%\bibliography{bibthese2}

\end{document}